\documentclass{ws-ijtaf}

\usepackage{graphics}
\usepackage{psfrag}
\usepackage{amsmath}
\usepackage{amssymb}

\begin{document}

\markboth{Shivanian, Lopez-Ruiz}
{A New Model for Ideal Gases}

\catchline{}{}{}{}{}

\title{A NEW MODEL FOR IDEAL GASES. \\ DECAY TO THE MAXWELLIAN DISTRIBUTION}

\author{ELYAS SHIVANIAN}
\address{Dept. of Mathematics, Imam Khomeini International University, \\
Qazvin, 34149-16818, Iran \\
\email{shivanian@ikiu.ac.ir}}

\author{RICARDO LOPEZ-RUIZ}
\address{DIIS \& BIFI, University of Zaragoza, \\
Zaragoza, 50009, Spain \\
\email{rilopez@unizar.es}}

\maketitle

\begin{history}
\end{history}

\begin{abstract}
In this work, a new model in kinetic gas theory for deriving the Maxwellian Velocity Distribution (MVD)
is proposed. We construct an operator that governs the discrete time evolution of the velocity 
distribution. This operator, which conserves the momentum and the energy of the ideal gas,
has the MVD as a fixed point. Moreover, for any initial out-of-equilibrium
velocity distribution, it is shown that the gas decays to the equilibrium distribution, 
that is, to the MVD.
\end{abstract}



\section{Introduction}
\label{intro}

The Boltzmann-Gibbs distribution (BGD) and the Maxwellian velocity distribution (MVD) 
are two fundamental distributions in statistical physics. The BGD represents the 
distribution of the states of a system in equilibrium and the MVD represents the 
distribution of velocities in an ideal gas in equilibrium.
Although they can be directly derived from the equilibrium statistical theory, nowadays
there was not an unified framework showing that both distributions can be seen
as fixed points of two different nonlinear operators acting on the space
of distributions. Evidently, it must be cited the Boltzmann theory \cite{boltzmann} 
as a successful attempt in this direction explaining the MVD from a microscopic 
point of view applied to kinetic gas theory. 

A scheme inspired in economic systems that has recently been proposed\cite{lopezruiz2011}
to explain the attractivity, and then the ubiquity, of BGD reads as follows: 
Let $p(m) {\mathrm d}m$ denote the PDF ({\it probability density function}) of money in
a multi-agent economic system, i.e. the probability of finding an agent of the ensemble
with money between $m$ and $m + {\mathrm d}m$.
Consider now the discrete time evolution of an initial money distribution $p_0(m)$ 
at each time step $n$ under the action of an operator $\cal T$, which represents the 
average effect on the system of many random binary interactions (with number of the order 
of the system size) between pairs of agents exchanging their money.
Thus, the system evolves from time $n$ to time $n+1$ to asymptotically
reach the equilibrium wealth distribution $p_f(m)$, i.e.
$$
\lim_{n\rightarrow\infty} {\cal T}^n \left(p_0(m)\right) \rightarrow 
p_f(x)=BGD=\delta e^{-\delta x} \hskip 0.5cm with \hskip 5mm \delta=<p_0>^{-1}\,.
$$
In this case, $p_f(m)$ is the exponential distribution (BGD) with the same
average wealth $<p_f>=\delta^{-1}$ than the initial one $<p_0>$,
due to the local and total money conservation\cite{yakoven2009}. The mathematical properties
of operator ${\cal T}$ have been disclosed in Ref.\cite{lopez2011}. 
Hence, this framework not only puts in evidence that the BGD is the equilibrium distribution
if not that in this case the BGD is asymptotically reached independently of the initial wealth distribution
given to the system, a point of view that to date was possibly lacking in 
the literature.

In this work, we extend this perspective to another problem of the same statistical nature.
Our goal is to explain the ubiquity of the MVD in ideal gases\cite{maxwell}. 
In the next section, we construct an operator $T$ in the space of velocity distributions 
in order to explain the decay of any initial velocity distribution to the MVD. 
Then, in section \ref{sec:2} we proceed to prove the dynamical properties 
of this operator $T$, in particular that the MVD is the fixed point of the system where 
the dynamics asymptotically evolves. In section \ref{sec:3}, several examples showing 
this dissipative behavior are depicted. Finally, our conclusions are given.

\section{The new model for ideal gases. Operator T}
\label{sec:1}

Consider an ideal gas with particles of unity mass in the three-dimensional ($3D$) space. 
As long as there is not a privileged direction in the equilibrium, we can take any direction
in the space and to study the discrete time evolution of the velocity distribution in that direction.
Let us call this axis $U$. We can complete a Cartesian system with two additional orthogonal 
axis $V,W$. If $p_n(u){\mathrm d}u$ represents the probability of finding 
a particle of the gas with velocity component in the direction $U$ comprised between 
$u$ and $u + {\mathrm d}u$ at time $n$, then the probability to have at this time $n$
a particle with a $3D$ velocity $(u,v,w)$ will be $p_n(u)p_n(v)p_n(w)$. 
The particles of the gas collide between them, and after a number of interactions
of the order of system size, a new velocity distribution is attained at time $n+1$. 
Concerning the interaction of particles with the bulk
of the gas, we make two simplistic and realistic assumptions in order to obtain
the probability of having a velocity $x$ in the direction $U$ at time $n+1$:
(1) Only those particles with an energy bigger than $x^2$ at time $n$ can contribute 
to this velocity $x$ in the direction $U$, that is, all those particles whose velocities 
$(u,v,w)$ verify $u^2+v^2+w^2\ge x^2$; (2) The new velocities after collisions are equally 
distributed in their permitted ranges, that is, 
particles with velocity $(u,v,w)$ can generate maximal velocities
$\pm U_{max}=\pm\sqrt{u^2+v^2+w^2}$, then the allowed range of velocities $[-U_{max},U_{max}]$
measures $2|U_{max}|$, and the contributing probability of these particles to the velocity $x$
will be $p_n(u)p_n(v)p_n(w)/(2|U_{max}|)$. Taking all together we finally get the expression 
for the evolution operator $T$. This is: 
$$
p_{n+1}(x)=Tp_n(x) = \int\int\int_{u^2+v^2+w^2\ge x^2}\,{p_n(u)p_n(v)p_n(w)\over 2\sqrt{u^2+v^2+w^2}}
\; {\mathrm d}u{\mathrm d}v{\mathrm d}w\,.
$$

Let us remark that we have not made any supposition about the type of interactions or collisions
between the particles and, in some way, the equivalent of the Boltzmann hypothesis of {\it molecular
chaos} would be the two simplistic assumptions we have stated on the interaction of particles with
the bulk of the gas. But now a more clear and understandable framework
than those usually presented in the literature appears on the scene. 
In fact, the operator $T$ conserves in time
the energy and the null momentum of the gas. Moreover, for any initial velocity
distribution, the system tends towards its equilibrium, i.e. towards the MVD. 
This means that
$$
\lim_{n\rightarrow\infty} T^n \left(p_0(x)\right) \rightarrow p_f(x)=MVD\;(1D\;case)\,.
$$
Let us proceed to show all these properties in the next section.

\section{Decay to Maxwellian distribution. Properties of $T$}
\label{sec:2}

First, in order to set up the adequate mathematical framework,
we provide the following definitions.

\begin{definition}
We introduce the space $L_1^+$ of positive functions (one-dimensional velocity
distributions) in the real axis,
$$
L_1^+(\Re)=\lbrace p(x):\Re\to \Re^+\cup\lbrace 0\rbrace,
\hskip 2mm \vert\vert p\vert\vert<+\infty\rbrace,
$$
with norm $\vert\vert\cdot\vert\vert$ defined by
$$
\vert\vert p\vert\vert=\int_{-\infty}^{+\infty} p(x) dx.
$$
Therefore for each $p\in L_1^+(\Re)$, we have $\vert\vert p\vert\vert\ge 0$. 
\end{definition}

\begin{definition}
\label{def-mean1}
For each $x\in\Re$, $p\in L_1^+(\Re)$, we define the operator T as follows
$$
Tp(x) = \int\int\int_{u^2+v^2+w^2\ge x^2}\,{p(u)p(v)p(w)\over 2\sqrt{u^2+v^2+w^2}}
\; {\mathrm d}u{\mathrm d}v{\mathrm d}w\,.
$$
Therefore we have $Tp(x)\ge 0$ for each $x\in\Re$, and so $\vert\vert Tp\vert\vert\ge 0$.
\end{definition}

\begin{remark}
The following relations hold:
$$
\int_{-\infty}^{+\infty}\left(\int_{u^2\ge x^2}(\cdot){\mathrm d}u\right){\mathrm d}x =
\int_{-\infty}^{+\infty}\left(\int_{-\vert u\vert}^{+\vert u\vert}(\cdot){\mathrm d}x\right){\mathrm d}u \,,
$$
$$
\int_{-\infty}^{+\infty}\left(\int\int_{u^2+v^2\ge x^2}(\cdot){\mathrm d}u{\mathrm d}v\right){\mathrm d}x =
\int_{-\infty}^{+\infty}\int_{-\infty}^{+\infty}\left(\int_{-\sqrt{u^2+v^2}}^{+\sqrt{u^2+v^2}}(\cdot)
{\mathrm d}x\right){\mathrm d}u{\mathrm d}v \,,
$$
$$
\int_{-\infty}^{+\infty}\left(\int\int\int_{u^2+v^2+w^2\ge x^2}(\cdot){\mathrm d}u{\mathrm d}v{\mathrm d}w\right)
{\mathrm d}x = \hskip 4.5cm
$$
$$
\hskip 4.5cm\int_{-\infty}^{+\infty}\int_{-\infty}^{+\infty}\int_{-\infty}^{+\infty}
\left(\int_{-\sqrt{u^2+v^2+w^2}}^{+\sqrt{u^2+v^2+w^2}}(\cdot){\mathrm d}x\right)
{\mathrm d}u{\mathrm d}v{\mathrm d}w \,.
$$
\end{remark}

\begin{theorem}
For any $p\in L_1^+(\Re)$, we have $||Tp||=||p||^3$.
\end{theorem}

\begin{proof}
Suppose that $p\in L_1^+(\Re)$ then
$$
||Tp||=\int_{-\infty}^{+\infty}Tp(x)dx= \hskip 3cm
$$
$$\hskip 1cm
=\int_{-\infty}^{+\infty}\left(\int\int\int_{u^2+v^2+w^2\ge x^2}
{p(u)p(v)p(w)\over 2\sqrt{u^2+v^2+w^2}}\,{\mathrm d}u{\mathrm d}v{\mathrm d}w\right){\mathrm d}x = 
$$
$$\hskip 1cm 
=\int_{-\infty}^{+\infty}\int_{-\infty}^{+\infty}\int_{-\infty}^{+\infty}
\left(\int_{-\sqrt{u^2+v^2+w^2}}^{+\sqrt{u^2+v^2+w^2}}{p(u)p(v)p(w)\over 2\sqrt{u^2+v^2+w^2}}
\,{\mathrm d}x\right){\mathrm d}u{\mathrm d}v{\mathrm d}w=
$$
$$ 
=\int_{-\infty}^{+\infty}\int_{-\infty}^{+\infty}\int_{-\infty}^{+\infty}
p(u)p(v)p(w){\mathrm d}u{\mathrm d}v{\mathrm d}w = \hskip 3cm
$$
$$
=\int_{-\infty}^{+\infty}p(u){\mathrm d}u\int_{-\infty}^{+\infty}p(v){\mathrm d}v
\int_{-\infty}^{+\infty}p(w){\mathrm d}w = ||p||^3 \,. \hskip 2cm
$$
\end{proof}

\begin{corollary} 
It is clear that $(Tp)(x)\ge 0$ for $x\in\Re$ and $Tp\in L_1^+(\Re)$. 
Also, it is straightforward to see that $Tp$ is a continuous 
function, i.e. $Tp\in\mathbb{C}(\Re)$. 
\end{corollary}

\begin{corollary}
Consider the set $B\subset L_1^+(\Re)$ of PDF (Probability Density Functions), i.e.
$B=\{p\in L_1^+(\Re): ||p||=1\}$. It holds that if $p\in B$ then $Tp\in B$. Clearly,
if $||p||=1$ then $||Tp||=1$, i.e. the action of $T$ on $B$ conserves the 
number of particles in the gas.
\end{corollary}

\begin{corollary}
Consider the following recursion:
$$
p_n(x)=Tp_{n-1}(x)  \hskip 1cm with \hskip 1cm p_0\in L_1^+(\Re)\,.
$$
Since $||Tp||=||p||^3$ therefore there are three possibilities depending
on $p_0$:
\begin{itemize}
\item If $||p_0||<1$, then $\lim_{n\to\infty} ||p_n(x)||=0$.

\item If $||p_0||>1$, then $\lim_{n\to\infty} ||p_n(x)||=+\infty$.

\item If $||p_0||=1$, then $||p_n(x)||=1$ $\forall n$, i.e $p_n\in B$ $\forall n$.
\end{itemize}
\end{corollary}

Despite our interest resides specifically in the PDFs set, i.e. in $B$,
some of the next theorems are also valid for  generic elements of $L_1^+$.

\begin{theorem}
The mean value of the velocity in the recursion $p_n=T^np_0$ is conserved in time. 
In fact, it is null for all $n$:
$$
<x,Tp>=<x,T^2p>=<x,T^3p>=\cdots=<x,T^np>=\cdots=0 \,,
$$
where
$$
<f,g>=\int_{-\infty}^{+\infty}f(x)g(x){\mathrm d}x\,.
$$
It means that the zero total momentum of the gas 
is conserved in its time evolution under the action of $T$.
\end{theorem}

\begin{proof}
Assume that $p\in B$ then 
$$
<x,Tp>=\int_{-\infty}^{+\infty}xTp(x)dx= \hskip 3cm
$$
$$\hskip 1cm
=\int_{-\infty}^{+\infty}x \left(\int\int\int_{u^2+v^2+w^2\ge x^2}
{p(u)p(v)p(w)\over 2\sqrt{u^2+v^2+w^2}}\,{\mathrm d}u{\mathrm d}v{\mathrm d}w\right){\mathrm d}x = 
$$
$$\hskip 1cm 
=\int_{-\infty}^{+\infty}\int_{-\infty}^{+\infty}\int_{-\infty}^{+\infty}
{p(u)p(v)p(w)\over 2\sqrt{u^2+v^2+w^2}}
\left(\int_{-\sqrt{u^2+v^2+w^2}}^{+\sqrt{u^2+v^2+w^2}} x\,{\mathrm d}x\right)
{\mathrm d}u{\mathrm d}v{\mathrm d}w=
$$
$$\hskip 1cm 
=\int_{-\infty}^{+\infty}\int_{-\infty}^{+\infty}\int_{-\infty}^{+\infty}
{p(u)p(v)p(w)\over 2\sqrt{u^2+v^2+w^2}}
\left[{x^2\over 2}\right]_{-\sqrt{u^2+v^2+w^2}}^{+\sqrt{u^2+v^2+w^2}} 
{\mathrm d}u{\mathrm d}v{\mathrm d}w = 0 \,.
$$
Now, we know that if $p\in B$  then $Tp\in B$, therefore
$<x,T^np>=0$ $\forall n$.
\end{proof}

\begin{theorem}
The total energy of the gas is conserved in time. 
In other words, for every $p\in B$ we have
$$
<x^2,p>=<x^2,Tp>=<x^2,T^2p>=<x^2,T^3p>=\cdots=<x^2,T^np>=\cdots \,.
$$
\end{theorem}

\begin{proof}
Assume that $p\in B$ then 
$$
<x^2,Tp>=\int_{-\infty}^{+\infty}x^2Tp(x)dx= \hskip 3cm
$$
$$\hskip 1cm
=\int_{-\infty}^{+\infty}x^2 \left(\int\int\int_{u^2+v^2+w^2\ge x^2}
{p(u)p(v)p(w)\over 2\sqrt{u^2+v^2+w^2}}\,{\mathrm d}u{\mathrm d}v{\mathrm d}w\right){\mathrm d}x = 
$$
$$\hskip 1cm 
=\int_{-\infty}^{+\infty}\int_{-\infty}^{+\infty}\int_{-\infty}^{+\infty}
{p(u)p(v)p(w)\over 2\sqrt{u^2+v^2+w^2}}
\left(\int_{-\sqrt{u^2+v^2+w^2}}^{+\sqrt{u^2+v^2+w^2}} x^2\,{\mathrm d}x\right)
{\mathrm d}u{\mathrm d}v{\mathrm d}w=
$$
$$\hskip 1cm 
=\int_{-\infty}^{+\infty}\int_{-\infty}^{+\infty}\int_{-\infty}^{+\infty}
{p(u)p(v)p(w)\over 2\sqrt{u^2+v^2+w^2}}
\left[{x^3\over 3}\right]_{-\sqrt{u^2+v^2+w^2}}^{+\sqrt{u^2+v^2+w^2}} 
{\mathrm d}u{\mathrm d}v{\mathrm d}w  \,.
$$
$$\hskip 1cm 
=\int_{-\infty}^{+\infty}\int_{-\infty}^{+\infty}\int_{-\infty}^{+\infty}
{p(u)p(v)p(w)\over 2\sqrt{u^2+v^2+w^2}}
\,{2\over 3}\,(u^2+v^2+w^2)^{2\over 3}{\mathrm d}u{\mathrm d}v{\mathrm d}w =
$$
$$\hskip 1cm 
=\int_{-\infty}^{+\infty}\int_{-\infty}^{+\infty}\int_{-\infty}^{+\infty}
\,{1\over 3}\,(u^2+v^2+w^2)\,p(u)p(v)p(w){\mathrm d}u{\mathrm d}v{\mathrm d}w =
$$
$$\hskip 1cm 
={1\over 3}\,\int_{-\infty}^{+\infty}u^2p(u){\mathrm d}u +
{1\over 3}\,\int_{-\infty}^{+\infty}v^2p(v){\mathrm d}v +
{1\over 3}\,\int_{-\infty}^{+\infty}w^2p(w){\mathrm d}w =
$$
$$
=\int_{-\infty}^{+\infty}x^2p(x){\mathrm d}x = <x^2,p>. 
$$
Now, we know that if $p\in B$  then $Tp\in B$, therefore
$<x^2,T^np>=<x^2,p>$ $\forall n$.
\end{proof}

\begin{theorem}
The one-parametric family of PDFs $p_{\alpha}(x)=\sqrt{\alpha\over\pi}e^{-\alpha x^2}$,
$\alpha\ge 0$, are fixed points of the operator $T$. In other words, $Tp_{\alpha}=p_{\alpha}$.
\end{theorem}

\begin{proof}
When $\alpha=0$ then $p_{\alpha}=0$, which is clearly a fixed point of $T$.
Suppose now that $\alpha\not=0$,
$$\hskip 1cm
Tp_{\alpha}=\int\int\int_{u^2+v^2+w^2\ge x^2}{p_{\alpha}(u)p_{\alpha}(v)p_{\alpha}(w)
\over 2\sqrt{u^2+v^2+w^2}}\,{\mathrm d}u{\mathrm d}v{\mathrm d}w = 
$$
$$\hskip 2cm
=\int\int\int_{u^2+v^2+w^2\ge x^2}{{\alpha\over\pi}{\sqrt{\alpha\over\pi}}
e^{-\alpha(u^2+v^2+w^2)}\over 2\sqrt{u^2+v^2+w^2}}\,
{\mathrm d}u{\mathrm d}v{\mathrm d}w.
$$
Transforming the integral region to spherical coordinates by the change of variables
$$
u=r\sin\psi\cos\theta, v=r\sin\psi\sin\theta, w=r\cos\psi,
$$
the proof becomes straightforward:
$$
Tp_{\alpha} = \int_{|x|}^{+\infty}\int_0^{\pi}\int_0^{2\pi}
{\alpha\over 2\pi}\,\sqrt{\alpha\over\pi}\,e^{-\alpha r^2}r\sin\psi 
{\mathrm d}\theta{\mathrm d}\psi{\mathrm d}r=
$$
$$
=\int_{|x|}^{+\infty}\int_0^{\pi}
\alpha\,\sqrt{\alpha\over\pi}\,e^{-\alpha r^2}r\sin\psi{\mathrm d}\psi{\mathrm d}r
=\int_{|x|}^{+\infty}
2\alpha\,\sqrt{\alpha\over\pi}\,re^{-\alpha r^2}{\mathrm d}r=
$$
$$
=\sqrt{\alpha\over\pi}\,\left[-e^{-\alpha r^2}\right]_{|x|}^{+\infty}
=\sqrt{\alpha\over\pi}\,e^{-\alpha x^2}=p_{\alpha}\,.
$$
\end{proof}

\begin{theorem}
The operator $T$ is Lipschitz continuous in $B$ with Lipschitz constant $\delta \leq 3$.
\end{theorem}

\begin{proof}
Suppose $p(x), q(x)\in B$, then
$$
||Tp-Tq||= \int_{-\infty}^{+\infty}\left|\int\int\int_{u^2+v^2+w^2\ge x^2}
{p(u)p(v)p(w)-q(u)q(v)q(w)\over 2\sqrt{u^2+v^2+w^2}}\,
{\mathrm d}u{\mathrm d}v{\mathrm d}w\right|{\mathrm d}x = 
$$
$$
\le \int_{-\infty}^{+\infty}\int\int\int_{u^2+v^2+w^2\ge x^2}
{|p(u)p(v)p(w)-q(u)q(v)q(w)|\over 2\sqrt{u^2+v^2+w^2}}\,
{\mathrm d}u{\mathrm d}v{\mathrm d}w{\mathrm d}x = 
$$
$$ 
=\int_{-\infty}^{+\infty}\int_{-\infty}^{+\infty}\int_{-\infty}^{+\infty}
\left(\int_{-\sqrt{u^2+v^2+w^2}}^{+\sqrt{u^2+v^2+w^2}}
{|p(u)p(v)p(w)-q(u)q(v)q(w)|\over 2\sqrt{u^2+v^2+w^2}}\,{\mathrm d}x\right)
{\mathrm d}u{\mathrm d}v{\mathrm d}w=
$$
$$
=\int_{-\infty}^{+\infty}\int_{-\infty}^{+\infty}\int_{-\infty}^{+\infty}
|p(u)p(v)p(w)-q(u)q(v)q(w)|\,{\mathrm d}u{\mathrm d}v{\mathrm d}w=
$$
$$
=\int_{-\infty}^{+\infty}\int_{-\infty}^{+\infty}\int_{-\infty}^{+\infty}
|p(u)p(v)p(w)-p(u)p(v)q(w)+p(u)p(v)q(w)- \hskip 1.5cm
$$
$$\hskip 3.2cm
-p(u)q(v)q(w)+p(u)q(v)q(w)-q(u)q(v)q(w)|
\,{\mathrm d}u{\mathrm d}v{\mathrm d}w=
$$
$$
\le\int_{-\infty}^{+\infty}p(u){\mathrm d}u \int_{-\infty}^{+\infty}p(v){\mathrm d}v
\int_{-\infty}^{+\infty} |p(w)-q(w)|{\mathrm d}w+
$$
$$\hskip 0.5cm
+\int_{-\infty}^{+\infty}p(u){\mathrm d}u \int_{-\infty}^{+\infty}q(w){\mathrm d}w
\int_{-\infty}^{+\infty} |p(v)-q(v)|{\mathrm d}v+
$$
$$\hskip 0.7 cm
+\int_{-\infty}^{+\infty}q(v){\mathrm d}v \int_{-\infty}^{+\infty}q(w){\mathrm d}w
\int_{-\infty}^{+\infty} |p(u)-q(u)|{\mathrm d}u =
$$
$$
= 3\,\int_{-\infty}^{+\infty} |p(x)-q(x)|\,{\mathrm d}x = 3\,||p-q||\,.
$$
\end{proof}

\begin{theorem}
Suppose that $\lim_{n\rightarrow\infty} ||T^np(x)-\mu(x)||=0$, 
and $\mu(x)$ is a continuous function in $B$, then $\mu(x)$ is the fixed point
of the operator $T$ for the initial condition $p(x)\in B$.
\label{theor-T-limit}
\end{theorem}

\begin{proof}
First we prove that for each $\epsilon>0$ there exists $M$ so that $\forall n>M$:
$||T^np-T\mu||<\epsilon$. Since 
$\lim_{n\rightarrow\infty} ||T^np(x)-\mu(x)||=0$,
we can say that $\exists N$ such that $\forall n>N$: $||T^{n-1}p-\mu||<\epsilon/3$.
Now, we choose $M=N$. Then, due to the Lipschitz continuity of $T$ in $B$, we have
$$
\forall n>M:||T^np-T\mu||\leq 3||T^{n-1}p-\mu||<
3\cdot {\epsilon\over 3} = \epsilon\,.
$$
Now, by uniqueness of the limit, it implies that $T\mu=\mu$. Therefore,
for the initial condition $p(x)\in B$,
$\mu(x)$ is the fixed point of $T$.
\end{proof}

\vskip 0.3cm
{\bf Conjecture:} For any $p\in B$, with finite $<x^2,p>$ and
verifying $\lim_{n\rightarrow\infty} ||T^np(x)-\mu(x)||=0$,
the limit $\mu(x)$ is the fixed point 
$p_{\alpha}(x)=\sqrt{\alpha\over\pi}\,e^{-\alpha x^2}$,
with $\alpha=(2\,<x^2,p>)^{-1}$.
In physical terms, it means that for any initial velocity distribution of the gas, 
it decays to the Maxwellian distribution, which is just the fixed point of the dynamics.
Recalling that $<x^2,p>=k\tau$, with $k$ the Boltzmann constant and $\tau$ the temperature
of the gas, and introducing the mass $m$ of the particles, let us observe that the MVD is 
recovered in its $3D$ format: 
$$
MVD = p_{\alpha}(u)p_{\alpha}(v)p_{\alpha}(w)=\left(m\alpha\over\pi\right)^{3\over 2}\,
e^{-m\alpha (u^2+v^2+w^2)} \hskip 0.5cm with \hskip 5mm \alpha=(2k\tau)^{-1}.
$$

\section{Examples}
\label{sec:3}

In this section, we guess by simulation of many examples the truth of the above Conjecture, 
that is:
$$
\lim_{n\rightarrow\infty} \left|\left|{T}^np(x)-\sqrt{\alpha\over\pi}\,
e^{-\alpha x^2}\right|\right|=0 \hskip 0.5cm with \hskip 0.5cm \alpha=(2\,<x^2,p>)^{-1}\,.
$$
See the next examples.

\begin{figure}[h]
\begin{center}
\psfrag{B}{} \psfrag{A}{\large\scriptsize (a)}
\includegraphics[width=6cm,height=5cm]{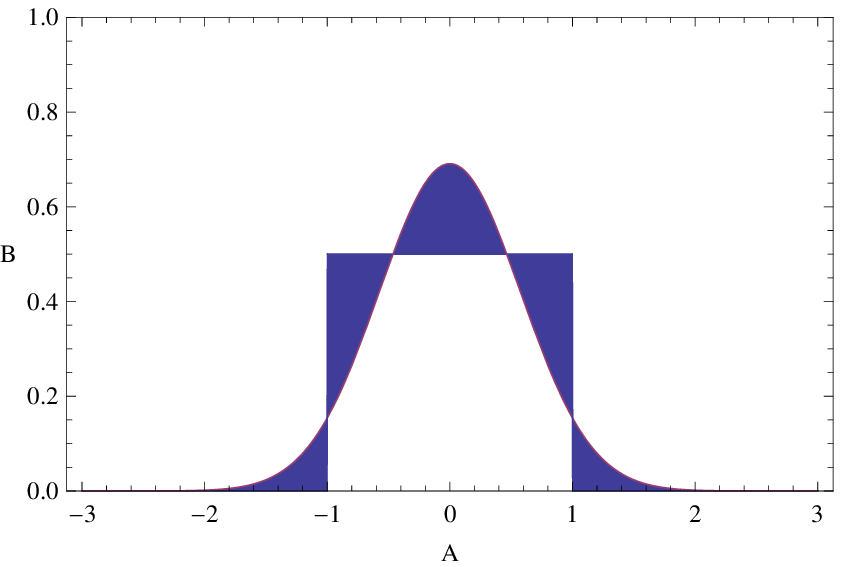} \hskip 2 mm
\psfrag{B}{} \psfrag{A}{\large\scriptsize (b)}
\includegraphics[width=6cm,height=5cm]{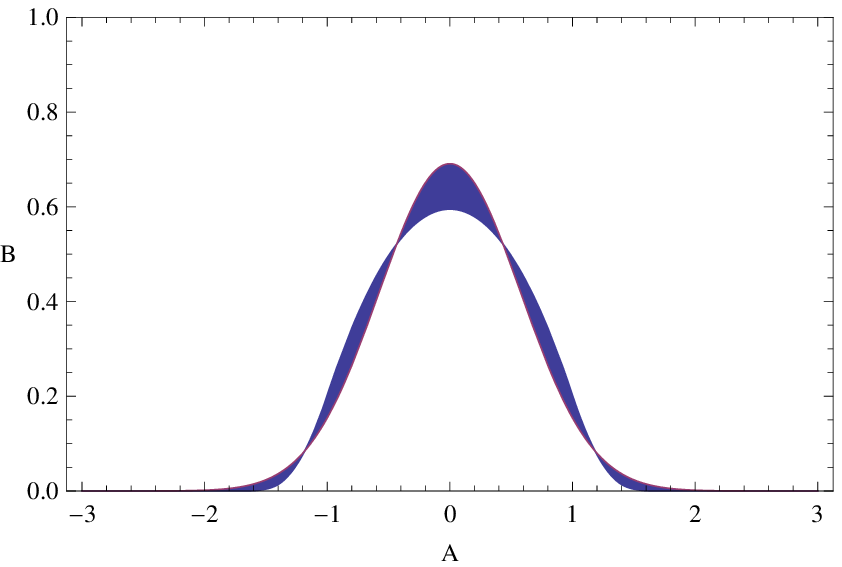} \hskip 2 mm
\caption{Example \ref{example-1}: (a) $p(x)$ and $\mu(x)$, (b) $Tp(x)$ and $\mu(x)$.}
\label{fig-1}
\end{center}
\end{figure}

\begin{example}
Take $p(x)$ a constant even function with mean value equal zero:
$$
p(x)=\left\{\begin{array}{cc}
{1\over 2} & \hskip 0.5cm -1<x<1\,, \\
0 & \hskip 0.5cm otherwise,
\end{array}\right.
$$
so $\alpha=(2\,<x^2,p>)^{-1}=3/2$. Therefore we expect that
$$
T^np(x)\stackrel{\small n\to\infty}{\longrightarrow}\mu(x)=\sqrt{3\over 2\pi}\,e^{-3x^2\over 2}.
$$
See the graphs of $p(x)$, $Tp(x)$ and $\mu(x)$ in Fig. \ref{fig-1}.
\label{example-1}
\end{example}

\begin{figure}[h]
\begin{center}
\psfrag{B}{} \psfrag{A}{\large\scriptsize (a)}
\includegraphics[width=6cm,height=5cm]{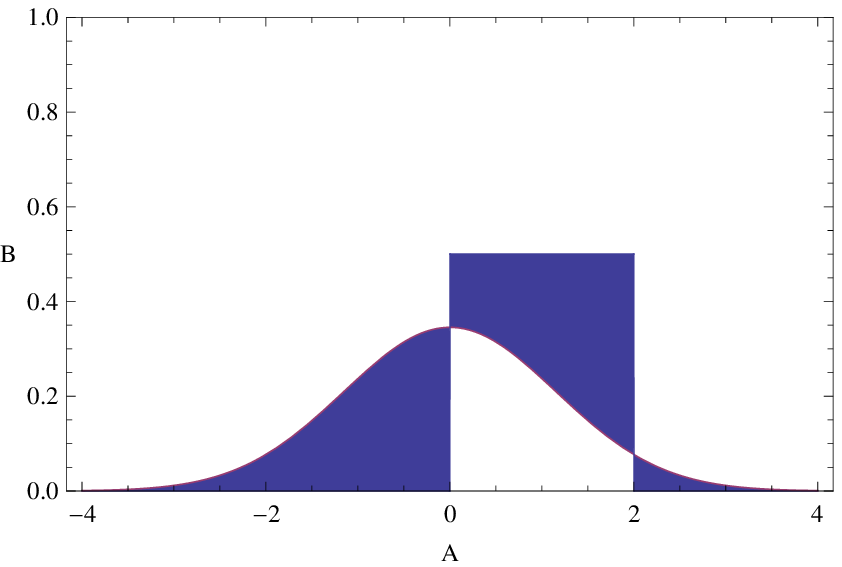} \hskip 2 mm
\psfrag{B}{} \psfrag{A}{\large\scriptsize (b)}
\includegraphics[width=6cm,height=5cm]{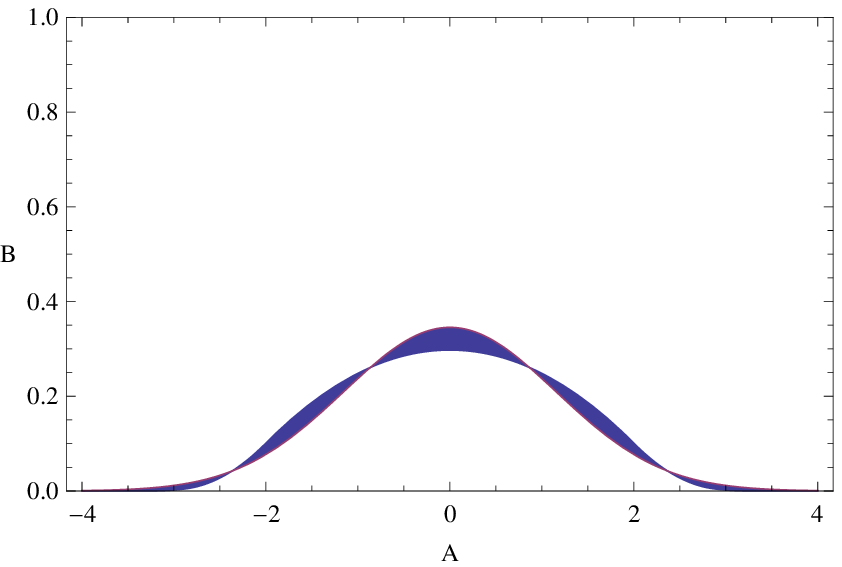} \hskip 2 mm
\caption{Example \ref{example-2}: (a) $p(x)$ and $\mu(x)$, (b) $Tp(x)$ and $\mu(x)$.}
\label{fig-2}
\end{center}
\end{figure}

\begin{example}
Take now $p(x)$ a constant even function with mean value different from zero:
$$
p(x)=\left\{\begin{array}{cc}
{1\over 2} & \hskip 0.5cm 0<x<2\,, \\
0 & \hskip 0.5cm otherwise,
\end{array}\right.
$$
so $\alpha=(2\,<x^2,p>)^{-1}=3/8$. Therefore we expect that
$$
T^np(x)\stackrel{\small n\to\infty}{\longrightarrow}\mu(x)=\sqrt{3\over 8\pi}\,e^{-3x^2\over 8}.
$$
See the graphs of $p(x)$, $Tp(x)$ and $\mu(x)$ in Fig. \ref{fig-2}.
\label{example-2}
\end{example}

\begin{figure}[h]
\begin{center}
\psfrag{B}{} \psfrag{A}{\large\scriptsize (a)}
\includegraphics[width=6cm,height=5cm]{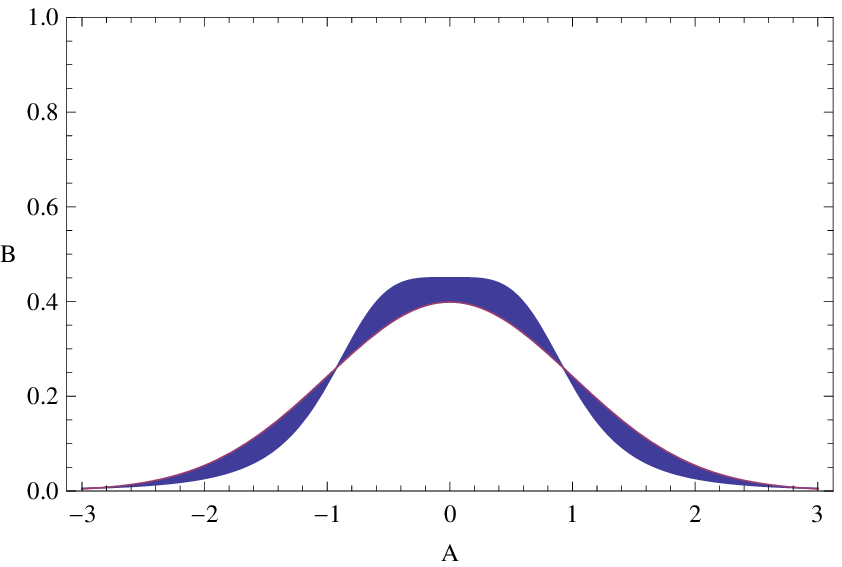} \hskip 2 mm
\psfrag{B}{} \psfrag{A}{\large\scriptsize (b)}
\includegraphics[width=6cm,height=5cm]{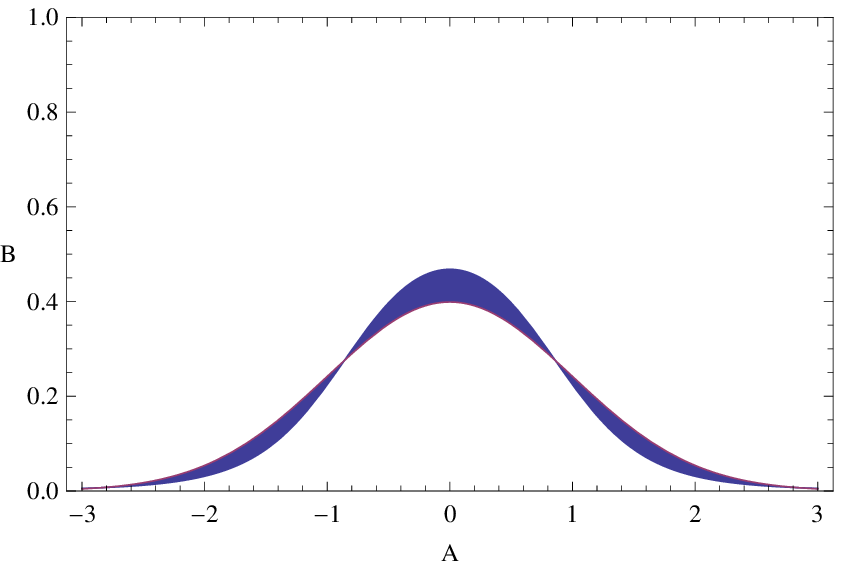} \hskip 2 mm
\caption{Example \ref{example-3}: (a) $p(x)$ and $\mu(x)$, (b) $Tp(x)$ and $\mu(x)$.}
\label{fig-3}
\end{center}
\end{figure}

\begin{example}
Assume $p(x)$ a non-constant even function with mean value equal zero:
$$
p(x)={\sqrt{2}\over \pi(1+x^4)}\,,
$$
so $\alpha=(2\,<x^2,p>)^{-1}=1/2$. Therefore we expect that
$$
T^np(x)\stackrel{\small n\to\infty}{\longrightarrow}\mu(x)=\sqrt{1\over 2\pi}\,e^{-x^2\over 2}.
$$
See the graphs of $p(x)$, $Tp(x)$ and $\mu(x)$ in Fig. \ref{fig-3}.
\label{example-3}
\end{example}

\begin{figure}[h]
\begin{center}
\psfrag{B}{} \psfrag{A}{\large\scriptsize (a)}
\includegraphics[width=6cm,height=5cm]{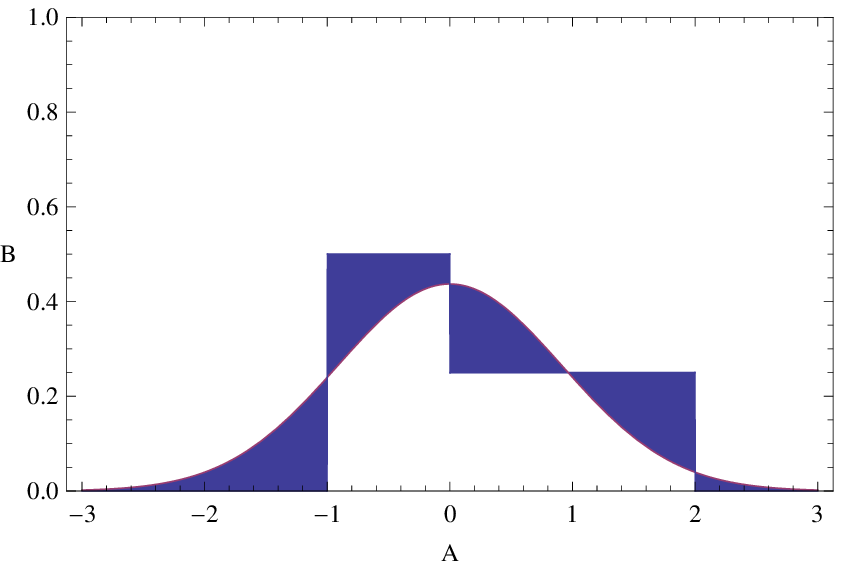} \hskip 2 mm
\psfrag{B}{} \psfrag{A}{\large\scriptsize (b)}
\includegraphics[width=6cm,height=5cm]{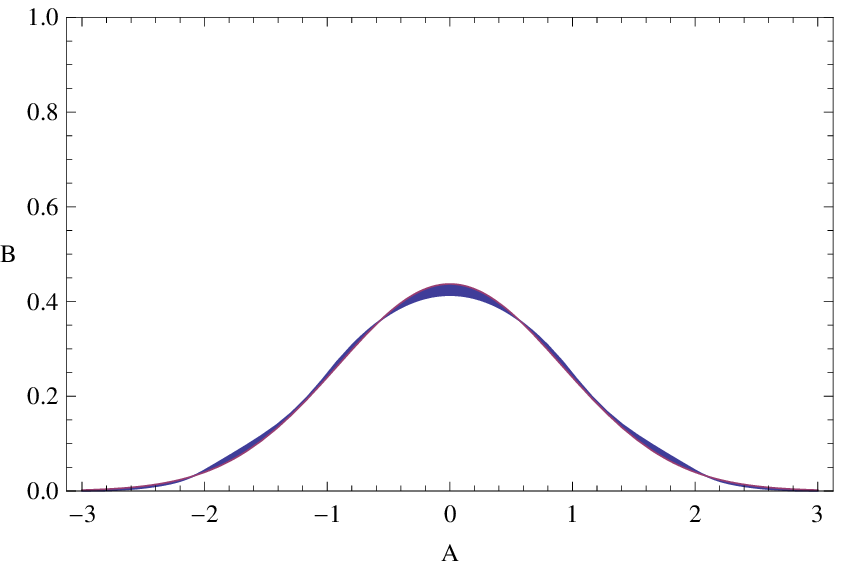} \hskip 2 mm
\caption{Example \ref{example-4}: (a) $p(x)$ and $\mu(x)$, (b) $Tp(x)$ and $\mu(x)$.}
\label{fig-4}
\end{center}
\end{figure}

\begin{example}
Take $p(x)$ a piecewise linear function with mean value equal zero:
$$
p(x)=\left\{\begin{array}{cc}
{1\over 2} & \hskip 0.5cm -1<x<0\,, \\
{1\over 4} & \hskip 0.7cm 0<x<2\,, \\
0 & \hskip 0.7cm otherwise\,, 
\end{array}\right.
$$
so $\alpha=(2\,<x^2,p>)^{-1}=3/5$. Therefore we expect that
$$
T^np(x)\stackrel{\small n\to\infty}{\longrightarrow}\mu(x)=\sqrt{3\over 5\pi}\,e^{-3x^2\over 5}.
$$
See the graphs of $p(x)$, $Tp(x)$ and $\mu(x)$ in Fig. \ref{fig-4}.
\label{example-4}
\end{example}

\begin{figure}[h]
\begin{center}
\psfrag{B}{} \psfrag{A}{\large\scriptsize (a)}
\includegraphics[width=6cm,height=5cm]{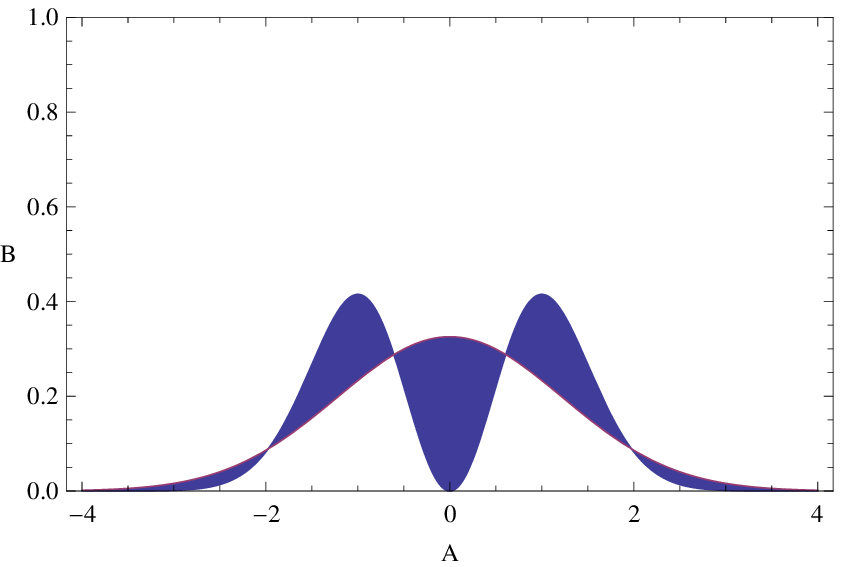} \hskip 2 mm
\psfrag{B}{} \psfrag{A}{\large\scriptsize (b)}
\includegraphics[width=6cm,height=5cm]{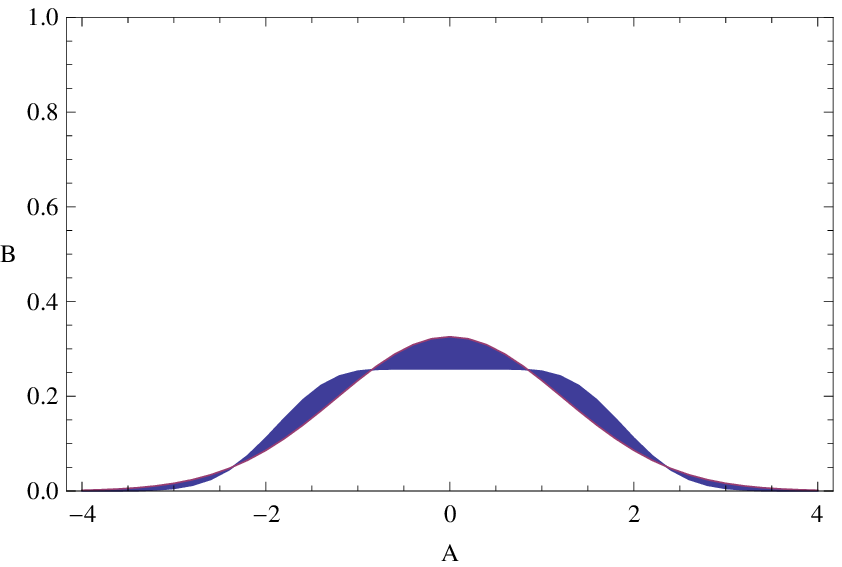} \hskip 2 mm
\caption{Example \ref{example-5}: (a) $p(x)$ and $\mu(x)$, (b) $Tp(x)$ and $\mu(x)$.}
\label{fig-5}
\end{center}
\end{figure}

\begin{example}
Finally, take $p(x)$ a more complex shaped function with mean value equal zero:
$$
p(x)={2\over\sqrt{\pi}}\,x^2e^{-x^2}\,, 
$$
so $\alpha=(2\,<x^2,p>)^{-1}=1/3$. Therefore we expect that
$$
T^np(x)\stackrel{\small n\to\infty}{\longrightarrow}\mu(x)=\sqrt{1\over 3\pi}\,e^{-x^2\over 3}.
$$
See the graphs of $p(x)$, $Tp(x)$ and $\mu(x)$ in Fig. \ref{fig-5}.
\label{example-5}
\end{example}

Let us observe that even if the initial $1D$ distribution $p(u)$ do not have the spherical
symmetry in the $3D$ format $p(u)p(v)p(w)$ (see for instance Example \ref{example-4}), 
the operator $T$ seems to recover this symmetry just after the first iteration. 
This suggests that in general the decay to the MVD is very fast, and after some 
few iterations it would be difficult to distinguish how far is the system of 
the asymptotic Maxwellian regime (MVD). In fact, we have also checked the increasing 
of entropy with time ({\it H-Theorem}), that is verified in all these examples. 
This is another important point now under study, 
that was also addressed in Boltzmann theory\cite{boltzmann} and 
that can be the subject of a future work for the present model.

\section{Conclusions}

In this work, a new model for ideal gases has been presented.
This model gives account of the decay of the velocity distribution of 
an ideal gas toward the Maxwellian distribution. The first step has been 
to construct, under two simplistic and realistic assumptions, a discrete 
time evolution operator in the space of velocity distributions.
The second step has consisted of showing the physical properties of this operator,
specifically the conservation in time of the total energy and zero momentum of the gas.
Third, the Maxwellian velocity distribution has been proved to be a fixed point
of the dynamics toward which the gas asymptotically evolves with time.
Finally, several examples, that display this behavior, have been explicitly given.

Let us conclude by saying that the type of models explained here have been successful 
in explaining the wealth distribution provoked by random markets and also the 
velocity distribution reached by ideal gases. We hope that more problems
coming from other contexts can be studied under this new framework
in the next future.

\section*{Acknowledgments}

R.L.-R. acknowledges financial support from  the Spanish research project
DGICYT-FIS2009-13364-C02-01.

\end{document}